\documentclass[12pt]{article}
\title{
Perturbative expansions in QCD and analytic properties of
$\alpha_s$}
\author{Yu.A.Simonov \\
State Research Center,\\ Institute of Theoretical and Experimental
Physics, Moscow, Russia}
\date{}
  \newcommand{\be}{\begin{equation}}
\newcommand{\ee}{\end{equation}}  
\def\fun#1#2{\lower3.6pt\vbox{\baselineskip0pt\lineskip.9pt
\ialign{$\mathsurround=0pt#1\hfil ##\hfil$\crcr#2\crcr\sim\crcr}}}
\newcommand{\vex}{\mbox{\boldmath${\rm x}$}}

\newcommand{\vek}{\mbox{\boldmath${\rm k}$}}
\newcommand{\vey}{\mbox{\boldmath${\rm y}$}}

\newcommand{\vez}{\mbox{\boldmath${\rm z}$}}

 \newcommand{\lan}{\langle}
 \newcommand{\ran}{\rangle}
\begin{document}

\maketitle

\begin{abstract}
It is shown that analytic  properties of standard QCD perturbation
theory  contradict known spectral properties,and contain in
particular IR generated ghost poles and cuts. As an outcome the
rigorous background perturbation theory is developed and its
analytic properties are shown to be in  agreement with general
requirements. In a limiting case of large $N_c$, when QCD
amplitudes contain only pole singularities, the strong coupling
constant $\alpha_s(Q)$ is shown to be also meromorfic function of
external momenta.

 Some simple models and examples are given when nonperturbative
 $\beta$ function and $\alpha_s(Q)$ can be  written explicitly.
 General form of amplitudes at large $N_c$ is given in the
 framework of background perturbation theory and its
 correspondence with standard perturbation theory at large momenta
 is demonstrated in the example of $e^+e^-$ annihilation. For
 time-like momenta the background coupling constant differs
 drastically from the standard one
  but the background series
  averaged over energy intervals has
   the same (AF) asymptotics at large momenta as in the standard perturbation theory.

 \end{abstract}

  \section{Introduction}

 The standard perturbation theory (SPT) in QCD is well developed
 both on the theoretical and on phenomenological levels \cite{1},\cite{2} ,
 and constitutes the major  and best understood part of QCD.
 Successful applications  of perturbative expansions for processes
 with large momenta are numerous and impressive.

 Howerever there are several basic difficulties in SPT (it is
 assumed that RG improvements such as  partial summation of large
 logarithms are automatically included in SPT), namely

 1) Analytic properties of SPT amplitudes do not correspond to the
 expected spectral behaviour. In particular, there appear ghost
 poles and cuts even in the Euclidean region of momenta, where one
 expects amplitudes to be holomorphic.

 These ghost singularities are due to analytic properties of RG
 improved $\alpha_s(Q)$ and are connected with IR divergence of
 $\alpha_s(Q)$. Also in the Minkowskian region of momenta
 $\alpha_s(Q)$ have logarithmic cuts which the physical amplitude
 should not possess,  and they are therefore artefact of SPT.

 In short, amplitudes computed in SPT have analytic behaviour
 which has nothing to do with physical thresholds and cuts due to
 creation of hadrons.  Therefore one may only speak about some
 duality relations between SPT amplitudes and physical amplitudes
 integrated over sufficiently large energy interval.

 2) Another basic defect of SPT  is the lack of convergence of
 standard perturbative series. There are arguments that the latter
 is an asymptotic series  \cite{3}, and in all cases it is not clear
 where the series should be cut off. (Some hints that the
 three-loop contribution deteriorates the physical results in the 1-1.5 GeV region  are
 contained in \cite{ 3}  and \cite{4}). Moreover, the Landau ghost poles
 give rise to  the appearence of the so-called IR renormalons
 \cite{5}, which make the sum of perturbative series undefined even
 in the Borel sense.

 The attempts to connect the IR renormalons with the
 nonperturbative contributions may have only qualitative character
 at best.
 Therefore strictly speaking the notion of the sum of SPT series
 has no definite meaning, and one may only hope that at large
 Euclidean momenta the first few terms of RG improved perturbative
 series describe the asymptotics of physical amplitudes with
reasonable accuracy.

 To improve the situation several approaches have been developed.
 First of all, it was understood long ago, that in addition to SPT
 also nonperturbative contributions should be taken into account.
 Technically the latter have been introduced as the local terms
 (condensates) in the OPE, and on this foundation the QCD sum
 rules have been introduced \cite{6}. In this way one can approach the
 low-energy region around 1-2 GeV and a lot of useful physical
 information have been obtained in this method for the last 20
 years \cite{7}.

 However in this method one does not solve both problems of SPT
 described above, and instead postulates that in addition to a few
 first SPT terms one  can add few power terms to imitate behaviour
 of physical amplitude not only in asymptotic region but also in
 the region of few GeV. In the lower energy region the OPE series
 has at least double divergence: due to explosion of $\alpha_s$
 near the Landau ghost pole and due to explosion of power terms
 $\frac{C_n}{Q^{2n}}$.

 It is important to note that the original OPE and QCD sum rules
 \cite{6} have been properly defined only in the Euclidean region, and
 the transition to  the time-like region is assumed to be done
 aposteriori, after all calculations are done in the Eulidean
 region.

 Definition of $\alpha_s(Q^2)$ in the Minkowskian region of $Q^2,
 Q^2<0$, is a problem by itself in SPT, since formal analytic
 continuation of two - and  more-loop expressions for
 $\alpha_s(Q^2)$ yield complex expressions violating explicit
 unitarity conditions in nonasymptotic region.

 Moreover as was stressed in \cite{8}, the invariant coupling $\bar
 g(Q)$ can be defined only in the space-like domain, and "inside
 the RG formalism there is no simple means for defining $\bar g (Q)$
 in the time-like region".

 New formalism for definition of $\alpha_s(Q^2)$ both in
 space-like and time-like region was developed in
 \cite{9} and a new method was suggested, called analytic perturbation
 theory (see \cite{8} for a review and further references),
  where $\alpha_s(Q^2)$ is forced
 to be analytic for $Q^2>0$ and a  special procedure is envisaged
  to continue $\alpha_s$ analytically into the time-like region.
The method allows to  get rid of the Landau ghost pole and cuts
and seriously improves convergence of the perturbative series.

 It is  clear however that  this is not unique way of analytic definition
  in the whole complex plane of $z$.

 In what follows we shall choose a completely different strategy.
 To simplify matter we shall consider below the limiting case of
 large $N_c$. In this case one can be sure that all physical
 amplitudes contain only poles as functions of external momenta \cite{10},
 and we shall require that perturbative expansion would reproduce
 the meromorphic properties of physical amplitudes, i.e.
 $\alpha_s(Q^2)$ have only singularities in the time-like region,
 and those are poles.

 To make this goal, one needs to use Background Perturbation
 Theory (BPT) instead of the standard one, and we shall derive
 rigorous formalism based on BPT developed in the seventies and
 eighties \cite{11} and generalized in \cite{12}-\cite{14} to include nonclassical
 background  and averaging over background fields.

 In this way one obtains a systematic formalism  which allows to
 express all terms of BPT through the irreducible correlators of
 background fields (integrals thereof) and renormalized coupling
 constant $\alpha_B$, which we shall denote $\alpha_B(s)$ to
 distinguish it from $\alpha_s(s)$ in SPT.

 It was argued in \cite{14} that $\alpha_B(s)$ and physical amplitudes
 satisfies the same RG equations, in particular
 Ovsyannikov-Callan-Simanzyk (OCS) equations, and the important
 difference in $\alpha_B(Q^2)$ from $\alpha_s(Q^2)$ lies in the
 character of its dependence on $Q^2$. It was demonstrated in \cite{12}-\cite{14}
 that $\alpha_B(Q^2)$ has the property of freezing, or saturation
 at small $Q^2$, i.e. it tends to a finite limit  $\alpha_B(0)$
 when $Q^2\to 0$ and has no singularities in  the whole Euclidean
 region $Q^2\geq 0$.

 This behaviour of $\alpha_B(Q^2)$ was tested repeatedly in
 $e^+e^-$ annihilation \cite{13, 14}, in fine structure of charmonium \cite{15} and
 bottomonioum levels \cite{16} and recently \cite{4} in comparison with
 accurate lattice data on small-distance behaviour of
 $\alpha_L(R)$. In all cases the same form of solution for
$ \alpha_B$ was used without free parameters which produced
results
 in a good agreement with experimental and lattice data.
In this way the phenomenon of freezing (saturation) was
demonstrated both theoretically as a result of confinement in
background fields \cite{12}, and phenomenologically in comparison
with lattice and experimental data.

With all that some important theoretical questions were not
answered, first of all what happens with $\alpha_B(s)$ in the
Minkowskian (time-like) region and what are analytic properties of
$\alpha_B(s)$ in the whole $s$-plane.

Secondly, what is connection of $\alpha_B(s)$ and other
nonperturbative definitions of $\alpha(s)$, e.g. lattice
definitions of $\alpha-\alpha_L(s)$.

In the present paper it is our purpose to study the problem of BPT
in the large $N_c$ limit, trying to elucidate several aspects.

Firstly, we formulate the foundations of BPT and rules for
calculation of perturbative series, starting from the purely
nonperturbative term.

Secondly, we derive the RG equations taking into account that
nonperturbative background is in general not classical solution
and subject to the vacuum averaging, which due to the 'tHooft
identity can be done independently of the perturbative field
averaging.

Thirdly, we find the most general solution of RG equations and in
particular the nonperturbative $\beta$ - function which has known
lowest term expansion, at the same time  $\alpha_B(Q^2)$ is
represented as a sum of pole terms.

To understand which kind of singularities  $\alpha_B(Q^2)$ may
have compatible with analytic properties of  physical amplitudes
we consider a simplified model and demonstrate that any finite
order perturbative expansion has additional singularities which
are eliminated when a partial summation of the perturbative series
is done.

As a next step we formulate  the generic perturbative expansion
for a physical amplitude, choosing an example of $e^+e^-$
annihilation into hadrons and write it as a sum over poles in the
time-like region with calculable coefficients.

At this point one may wonder how this meromorphic expansion is
connected to SPT expansion, where the perturbative series is in
powers of $ \alpha_s(Q^2), $ which contain logarithmic functions
of $Q^2$ and not meromorphic. The answer to this connection was
given in the analysis of the lowest term of $e^+e^-$ annihilation
- the hadronic part of the photon self-energy. The latter is a sum
over meson poles at large $N_c$ and has a proper logarithmic
behaviour $\log Q^2$ at large $Q^2$. We demonstrate that a similar
correspondence takes place also for higher terms of BPT, and
formulate conditions on the coefficients of meromorphic expansion
which are necessary to reproduce the known SPT expansion at large
$Q^2$.

Discussion of practical applications and comparison to other
approaches concludes the paper.

The exact structure of paper is as follows. In section 2 general
rules of BPT are given, and Green's functions for valence quarks,
gluons and hadrons are written  explicitly.

In section 3 the RG equations are written down in the case of most
general background and the point of RG scheme is discussed and a
particular solution of RG equations for $\beta$- function and
$\alpha_B(Q^2)$ is presented.

In section 4 a simple model is discussed where some BPT subseries
can be summed up explicitly.

In section 5   nonasymptotic  terms in the  spectral sums are
calculated and in section 6  behaviour of perturbtive series in
the region of time-like $Q^2$ is studied in detail.

Discussion  and outlook is contained in the concluding section.

\section{Background Perturbation
Theory for arbitrary background}

Gluon field plays in QCD two different roles:

i) gluons are propagating, and at small distances this process can
 be described perturbatively, leading in particular  to color Coulomb
 interaction between quarks (antiquarks);

 ii) gluons form a kind of condensate, which serves as a background
 for the propagating perturbative gluons and quarks. This background
 is Euclidean and ensures  phenomena  of  confinement and CSB.

 Correspondingly we shall separate  the total gluonic field $A_\mu$
 into perturbative part $a_\mu$ and nonperturbative (NP) background
 $B_\mu$:
 \be
 A_\mu=B_\mu+a_\mu
 \label{9}
 \ee

 There are many questions about this separation, which may be
 answered now only partially, e.g. what exactly is the criterion of
 separation. Possible answer is that perturbative fields $a_\mu$ get
 their dimension from distance (momentum), and   therefore all
 correlators of   fields $a_\mu$ (in absence of $B_\mu$) are singular
 and made of
 inverse powers of ($x-y)$ and logarithms, where enters the only
 dimensional parameter of perturbative QCD -- $\Lambda_{QCD}$.
 Therefore evidently any dimensionful constant, like hadronic masses
 or string tension cannot be obtained as a perturbation series. In
 contrast to that, NP fields  $B_\mu$ have mass dimension due to the
 violation of scale invariance
 which is  intrinsically  present in the
 gluodynamics Lagrangian. The origin of separation (\ref{9}) is
 clearly seen in the  solutions of nonlinear equations for field
 correlators \cite{17}:  a perturbative  solution of those leads  to
 singular power-like field correlator, whereas at large distances
 there is a selfconsistent solution  of the equations, decaying
 exponentially with distance with arbitrary mass scale, since
 equations in  \cite{17} are scale--invariant. Full solution
 including intermediate distances produces mixed
 perturbative--nonperturbative terms,  containing both inverse powers
 of distance and exponentials.  For these terms criterion of
 separation fails.

 One can avoid formally the question of separation principle (and of
 double counting) using t'Hooft identity \cite{13}, which allows to
 integrate in (1) independently over $B_\mu$ and $a_\mu$:
 \be
 Z=\frac{1}{N'}\int DB_\mu \eta (B) D\psi D\bar \psi  D a_\mu
 e^{L_{tot}}
 \label{10}
 \ee
 Here
 the  weight $\eta (B)$ is arbitrary and may be  taken as constant.

To define the perturbation theory  series in $ga_\mu$ one starts
from (\ref{9}) and  rewrites the Lagrangian as follows: $$
L_{tot}=L_{gf}+L_{gh}+L(B+a)= $$
\be
L_0+L_1+L_2+L_{int}+L_{gf}+L_{gh} \label{12} \ee where $L_i$ have
the form:
\begin{eqnarray}
L_2(a)&=&\frac{1}{2} a_{\nu}(\hat{D}^2_{\lambda}\delta_{\mu\nu} -
\hat{D}_{\mu}\hat{D}_{\nu} + ig \hat{F}_{\mu\nu}) a_{\mu}=
\nonumber \\ &=&\frac{1}{2}
a^c_{\nu}[D_{\lambda}^{ca}D_{\lambda}^{ad} \delta_{\mu\nu} -
D_{\mu}^{ca}D_{\nu}^{ad} - g~f^{cad}F^a_{\mu\nu}]a^d_{\mu}~~,
\label{13}
\end{eqnarray}
$$ D_{\lambda}^{ca} =
 \partial_{\lambda}\cdot \delta_{ca}+ g~f^{cba} B^b_{\lambda}
\equiv \hat{D}_{\lambda},~~ F^a_{\mu\nu} =\partial_\mu
B^a_\nu-\partial_\nu B^a_\mu + g f^{abc} B^b_\mu B^c_\nu $$ $$ L_0
= -\frac{1}{4}
 (F^a_{\mu \nu}(B))^2 ~;~~~ L_1=a^c_{\nu} D_{\mu}^{ca}(B) F^a_{\mu\nu}
$$
\be
L_{int} = -\frac{1}{2} (D_{\mu}(B)a_{\nu} -D_{\nu}(B)a_{\mu})^a
g~f^{abc} a_{\mu}^b a_{\nu}^c - \frac{1}{4} g^2 f^{abc} a_{\mu}^b
a_{\nu}^c f^{aef} a^e_{\mu} a^f_{\nu} \label{14} \ee

\begin{eqnarray}
L_{ghost}=-\theta^+_a (D_{\mu}(B)D_{\mu} (B+a))_{ab} \theta_b.
\label{15}
\end{eqnarray}

 It is convenient to prescribe to $B_{\mu}$, $a_{\mu}$ the
 following gauge transformations
 \be
 a_{\mu}\to U^+a_{\mu}U\label{16}
 \ee
 \be
 B_{\mu}\to U(B_{\mu}+\frac{i}{g}\partial_{\mu})U
\label{17} \ee
 and to impose on $a_{\mu}$ the background gauge condition
  \be
 G^a\equiv (D_{\mu}a_{\mu})^a=\partial_{\mu}a_{\mu}^a+g
 f^{abc}B^b_{\mu}a^c_{\mu}=0\label{18}
 \ee
 In this case the ghost field have to be
  introduced  as in (\ref{15}) and the gauge-fixing term is $L_{gf}
  =-\frac{1}{2\xi}(G^a)^2$
One
 can write the resulting partition function as
 \be
 Z=\frac{1}{N'}\int DB\eta (B)
 Z(J,B)\label{19}
 \ee
 where
\begin{equation}
Z(J,B) = \int D\psi D\bar \psi D a_\mu D\theta D\theta^+
e^{L_{tot}+ \int J_\mu a_\mu dx}\label{20}
\end{equation}

  We now can identify the propagator of $a_{\mu}$ from the quadratic terms
in Lagrangian $L_2(a) - \frac{1}{2\xi}(G^a)^2$
\begin{eqnarray}
G^{ab}_{\nu\mu} = [\hat{D}^2_{\lambda} \delta_{\mu \nu} -
\hat{D}_{\mu}\hat{D}_{\nu} + ig \hat{F}_{\mu\nu} +\frac{1}{\xi}
\hat{D}_{\nu} \hat{D}_{\mu}]^{-1}_{ab}\label{21}
\end{eqnarray}
It will be convenient sometimes to choose $\xi =1$ and end up with
the well-known
 form of propagator in -- what one would call --
the background Feynman gauge
\begin{eqnarray}
G^{ab}_{\nu\mu} = (\hat{D}^2_{\lambda}\cdot  \delta_{\mu \nu} -
2ig \hat{F}_{\mu\nu}) ^{-1}\label{22}
\end{eqnarray}

  Integration over ghost and gluon degrees of freedom in (\ref{20}) yields
  \be
  Z(J,B)=const (det W(B))^{-1/2}_{reg}[det
  (-D_{\mu}(B)D_{\mu}(B+a)]_{a=\frac{\delta}{\delta
  J}}\times\label{23}
  \ee
  $$
  \times \{ 1+\sum^{\infty}_{l=1}
  \frac{S_{int}}{l!}(a=\frac{\delta}{\delta J})\}
  exp (-\frac{1}{2}JW^{-1}J)\left|_{J_{\nu}=D_{\mu}(B)F_{\mu\nu}(B)}
\right.  $$
  where
  $W=G^{-1}$, and $G $is defined in (\ref{21}-\ref{22}).

  Let us mention the important property of the background Lagrangian
  (\ref{12}): under gauge tranformations the fields $a_{\mu},B_{\mu}$
  transform as in (\ref{16}-\ref{17}) and all terms of (\ref{12}), including the
  gauge fixing one $\frac{1}{2}(G^a)^2$ are gauge invariant. That was
  actually one of the aims put forward by G. 'tHooft in \cite{11}, and it
  has important consequences too: (i)  any amplitude in perturbative
  expansion in $ga_{\mu}$ of (\ref{20}) and (\ref{23}) corresponding to a
  generalized Feynman diagram, is separately gauge invariant (for
  colorless initial and final states of course).

  (ii)  Due to gauge invariance of all terms, the renormalization is
  specifically simple in the background field formalism \cite{11}, since
  the counterterms enter only in gauge--invariant combination, e.g.
  $F^2_{\mu\nu}$ and hence $Z$--factors $Z_g$  and $Z_A$ are
  connected. We shall exploit this fact in Sections 6 and 7.

  Let us turn now to the  term  $L_1$ in (\ref{12})
  $L_1=2 tr(a_{\nu}D_{\mu}F_{\mu\nu})$, which is usually missing in
 the standard background field formalism \cite{11}, since one  tacitly assumes
 there that the background $B_{\mu}$ is a classical solution,
 \be
 D_{\mu}(B)F_{\mu\nu}(B)=0.\label{24}
 \ee
 Here we do not impose the condition (\ref{24}) and consider
 any background, classical or purely quantum
 fluctuations. Let us estimate the influence of the
 vertex $L_1$. In general it leads to the shift of the
 current $J_{\mu}$ in the expression for the
 perturbative series (\ref{23}). Physically it means that at
 each point the background $B_{\mu}$ can generate a
 perturbative gluon via the vertex
 $(a_{\mu}D_{\nu}F_{\nu\mu})$, and this vertex is
 proportional to the degree of "nonclassicality" of
 $B_{\mu}$. For the quasiclassical vacuum, like the
 instanton model, the average value\\
 $<(D_{\mu}F_{\mu\nu})^2>$ over instanton  ensemble is
 less then $0(\rho^4/R^4)$ and is small ( at most of the
 order of few percents) while average of
 $<D_{\mu}F_{\mu\nu}>$ vanishes in the symmetric vacuum.
 All this is true provided the instanton  gas  stabilizes at small
 density.

 Let us estimate the effect of $L_1$ in the general
 quantum case. To this end we calculate as in [12] the
 contribution of $L_1$ to the gluon propagator.

If one denotes by $<>_a$ the integral $Da_{\mu}$ with the weight
$L_{tot}$ as in (\ref{20}), we obtain
\begin{eqnarray}
<a_{\mu_1}(x_1) a_{\mu_2}(x_2)>_a= G_{\mu_1\mu_2} (x_1, x_2)
+\Delta^B_{\mu_1\mu_2} \nonumber
\\ \Delta^B_{\mu_1 \mu_2} \equiv \int d^4y_1 d^4y_2
G_{\mu_1\nu_1}(x_1,y_1)D_{\rho}F_{\rho \nu_1}(y_1) D_{\lambda}
F_{\lambda\nu_2}(y_2) G_{\nu_2\mu_2} (y_2,x_2)\label{25}
\end{eqnarray}
The gluon Green's function $G_{\mu\nu}$ is given in (\ref{22}) and
depends on the background field $B_{\mu}$, as well as $D_{\mu}$
and $F_{\rho\lambda}$.
 To get a simple estimate of $\Delta$ we
replace $G_{\mu\nu}$ by the free Green's function
$G^{(0)}_{\mu\nu}$ and take into account that \begin{eqnarray}
<D_{\rho}F_{\rho \nu_1}(y_1) D_{\lambda}
F_{\lambda\nu_2}(y_2)>_B\Rightarrow \frac{\partial}{\partial
y_{1\rho}}\frac{\partial}{\partial y_{2\lambda}}
<F_{\rho\nu_1}(y_1)F_{\lambda\nu_2}(y_2)> \label{26}\end{eqnarray}
and for the latter we use the representation \cite{18}  in terms
of two independent Lorentz structures, $D(y_1-y_2)$ and
$D_1(y_1-y_2)$. The contribution of $D$  (that of $D_1$ is of
similar character) in the momentum space   is
\begin{eqnarray}
\Delta^B_{\mu_1\mu_2}(k) \sim
\frac{(k^2\delta_{\mu_1\mu_2}-k_{\mu_1}k_{\mu_2})}{k^4}D(k)\label{27}
\end{eqnarray}
where
\begin{eqnarray}
D(k) = \int d^4y e^{iky}D(y)\label{28}
\end{eqnarray}
Insertion in (\ref{28})of the exponential fall-off for $D(y)$
found in lattice calculations \cite{19} yields
 \begin{eqnarray}
  D(k) =
\frac{<F^2(0)>}{(N^2_c-1)}\cdot
\frac{\pi^2\mu}{(\mu^2+k^2)^{5/2}}~.\label{29}
\end{eqnarray}

With $\mu\approx 1 GeV,~~D(k)\approx 0.12$. Thus $\Delta^B(k)$ is
a soft correction fast decreasing with $k$ as $k^{-5}$ to the
perturbative gluon propagator.

\section{Renormalization group improvement of perturbative series
and analytic properties of $\alpha_s(Q^2)$}

The most important property of the Background method discussed in
the previous chapter, is the gauge invariance of the total
Lagrangian including  the gauge-fixing term $L_{gf}$.

As a consequence of this the counterterms in the renormalization
procedure have to be gauge invariant too and this fact establishes
connection between $Z$-factors of  charge and background field
\cite{11}, namely, if
\be
g_0=Z_g g,~~ B_\mu^{(0)}=Z_B^{1/2} B_\mu \label{3.1} \ee then
\be
Z_g Z_B^{1/2}=1. \label{3.2} \ee

This property is very important for the BPT and the Field
Correlator Method in general. Indeed the combinations entering in
the latter have the form

\be
\Delta_n \equiv  tr \lan g F_{\mu_1\nu_1}^{(B)} (1) \Phi(1,2) g
 F_{\mu_2\nu_2}^{(B)} (2) \Phi(2,3)... g  F_{\mu_n\nu_n}^{(B)} (n) \Phi(n,1)
 \ran_B
 \label{3.3}
 \ee
 where $F^{(B)}_{\mu_i\nu_i}$
 is made of the background
 field $B_\mu$ only, and parallel transporters $\Phi(i,k)$ making Eq.(\ref{3.3})
 gauge invariant depend also on $g B_\mu$. Hence all expression in
 (\ref{3.3}) contains only combinations $g B_\mu$, which
 according
 to (\ref{3.2}) are renorminvariant.

 Therefore in all RG relations, in particular in
 Ovsyannikov-Callan-Simanzyk (OCS) equations the  background fields (averaged
 as in $\Delta_n$ or nonaveraged) enter on the same ground as the
 external momenta, and hence all relations can be kept intact.

 E.g. OCS equations for some physical amplitude $\Gamma$ now look
 like \cite{14}.
 \be
 \left\{-\frac{\partial}{\partial\log \lambda} + \beta
 \frac{\partial}{\partial g^2} + m (\gamma_m-1)
 \frac{\partial}{\partial m}-\gamma_\Gamma\right\} \Gamma(\lambda
 p_1, ... \lambda p_N,\lambda^{2k} \Delta_k, g,\nu)=0\label{3.4}
 \ee
 where we have defined as usual
 \be
 \frac{d\alpha_s(\lambda)}{d\log
 \lambda}=\beta(\alpha_s)\label{3.5}
 \ee
 and similarly for $\gamma_m$, with boundary conditions
 \be
 \alpha_s(\lambda=1) = \alpha_s(\nu).\label{3.8}\ee

 In writing the r.h.s. of (\ref{3.8}) we tacitly assume that
 $\alpha_s$ depends also on all external parameters $\{p_i\},
 \{\Delta_k\}$ in addition to the dimensional regularization
 parameter $\nu$ (or its equivalent in other schemes $\mu$). The
 reason why external parameters are not usually written lies in
 the fact that RG equations (like (\ref{3.4}), (\ref{3.5})) define
 only dependence on  one scale parameter, e.g. $\lambda $ and
 through it on $\nu$ $(\mu)$, while dependence on external
 parameters is not fixed, and they may in principle enter
 $\alpha_s$ in arbitrary dimensionless combinations, unless
 additional information is obtained from perturbation expansion
 (BPT or SPT).

 Several words should be said about the choice of renormalization
 scheme for BPT.

 It is convenient to use the $\overline{MS}$ scheme, since also in
 the presence of background the modified Feynman amplitudes of BPT
 are diverging at small distances (large momenta) and, when these
 momenta are much larger than the average magnitude of the
 background fields, one can neglect the latter, e.g. using OPE one
 expects corrections of the form $\frac{g^2\lan
 F^2_{\mu\nu}\ran}{p^4}$ which do not alter $Z$ factors. In
 particular the difference $\Gamma-\Gamma_{pert}\equiv \Delta
 \Gamma$, where $\Gamma_{pert}$ is a usual sum of Feynman
 amplitudes without background, is UV convergent, so that all UV
 divergencies of $\Gamma$ are the same as in $\Gamma_{pert}$.

 Now we turn to the possible solutions of Eq. (\ref{3.5}). The SPT
 expansion for $\beta(\alpha_s)$ looks like
 \be
 \beta^{pert}(\alpha_s)=-\frac{b_0}{2\pi}
 \alpha^2_s-\frac{b_1}{4\pi^2}
 \alpha^3_s-\frac{b_2}{64\pi^3}\alpha^4_s-\frac{b_3\alpha_s^5}{(4\pi)^4}-...
 \label{s.2}\ee
 where the first coefficients are\footnote{Note that definition of
 our coefficients $b_i$ differ from that in \cite{1} and coincide with [30]}
\be
SU(3);~ n_f=0;~ b_0=11;~ b_1=51,~ b_2(\overline{MS}) =2857,~
b_3(\overline{MS})=58486. \label{s.4} \ee
\be
  SU(N_c);~ n_f=0;~ b_0=\frac{11}{3} N_c;~ b_1=\frac{17}{3} N^2_c.
  \label{s.5}
  \ee

  Following (\ref{s.2}) it is reasonable to look for
  $\beta(\alpha_s) $ in the form
  \be
  \beta(\alpha_s)=-\frac{b_0}{2\pi}
  \frac{\alpha_s^2}{[1-\frac{b_1}{2\pi b_0}
  \varphi'(\frac{1}{\alpha_s})]}\label{s.1}
  \ee
  where $\varphi(x)$ is an unknown function and $\varphi'(x)$ is
  the derivative in the argument $x$. Solving (\ref{s.1}) for
  $\alpha_s$ one obtains

   \be
   \alpha_s=
   \frac{4\pi}{b_0[\ln \mu^2C^2+\frac{2b_1}{ b_0^2}
   \varphi(\frac{1}{\alpha_s})]}\label{s.6}
   \ee
where we have replaced $\lambda\to \mu$,
 and  $C^2$ is an arbitrary function of external parameters of dimension (mass)$^{-2}$.

>From (\ref{s.2}) one concludes that the only condition on
$\varphi$ is the
 expansion in powers of $\alpha_s$
 \be
  \frac{1}{1-\frac{b_1}{2\pi b_0}
  \varphi'(\frac{1}{\alpha_s})}=1+\frac{b_1}{2\pi
  b_0}\alpha_s+\frac{b_2}{32\pi^2 b_0}\alpha^2_s+\frac{b_3\alpha^3_s}{128\pi^3
  b_0}.
  \label{s.7}\ee
One can rewrite (\ref{s.6}) in the form
 \be
  \chi(Q^2)=z-\frac{b_1}{2\pi b_0}\varphi(z)
  \label{s.7a}
  \ee
  where $z\equiv1/\alpha_s$
  and $\chi(Q^2)\equiv \frac{b_0}{4\pi} \ln \mu^2C^2$, and $Q^2$
  is the representative of the external parameters.

  As a simple example one can consider the self-energy photon
  function $\Pi(Q^2)$ entering in the $e^+e^-$ annihilation
  process, in which case $Q^2$ is the only external momentum,
  while background fields create masses of the bound state
  spectrum. In this case one can take the form appearing in the
  lowest  approximation of BPT \cite{14}, leading to the replacement
  \be
\left\{  \begin{array}{l}
\ln\mu^2C^2=\ln\frac{m^2}{\Lambda^2}+\psi\left(\frac{Q^2+M^2_0}{m^2}\right)\\
\Lambda^2=\nu^2 e^{-\frac{4\pi}{b_0} \alpha_s (\nu^2)}
\end{array}\right.
\label{s.8} \ee

Here $\psi(x)\equiv \frac{\Gamma'(x)}{\Gamma(x)}$ is the Euler
function, having only simple  poles at $x=-n$, $n=0,1,2,...$.

 This
example clearly shows that $\chi(Q^2)$ is a meromorphic function
of $Q^2$, and we must require that $\alpha_s$ be also a
meromorphic function of $Q^2$, analytic in the Euclidean region
$Re Q^2\geq 0$, since otherwise the BPT would violate the analytic
properties of amplitudes in the limit $N_c\to \infty$.

In this way we are lead to the conclusion that $\varphi(z)$ should
be a meromorphic function of $z=\frac{1}{\alpha_s}$, since in this
case the r.h.s. of Eq.(\ref{s.7}) is a meromorphic function of
$Q^2$ (meromorphic function of a meromorphic function of some
argument is again meromorphic). Thus one can equalize functions on
both sides of Eq.(\ref{s.7}) since both have the same analytic
properties.

We do not specify at this point the character of possible
singularities; the resulting physical amplitude  should have only
simple poles at bound state energies, while any finite series of
BPT may have poles of higher order and additional poles, as will
be seen in examples below.

To illustrate possible forms of $\varphi(z)$ let us consider one
example of meromorphic function, namely
\be
\varphi\left
(\frac{1}{\alpha_s}\right)=\psi\left(\frac{1}{\alpha_s}+\Delta\right),~~
\Delta= const. \label{s.9}\ee Perturbative expansion of $\psi$ has
the form
\be
\psi'(z)=\frac{1}{z}+\frac{1}{2z^2}+\frac{1}{6z^3}+...,~~
z\to\infty \label{s.10} \ee or applying it to (\ref{s.9}) one has
\be
\psi'\left(\frac{1}{\alpha_s}+\Delta\right)=\frac{\alpha_s}{1+\alpha_s\Delta}
+\frac{1}{2} \frac{\alpha^2_s}{(1+\alpha_s\Delta)^2}
+\frac{\alpha^3_s}{6(1+\alpha_s\Delta)^3}+...\label{s.11}\ee
Therefore one obtains $$ \frac{1}{1-\frac{b_1}{2\pi b_0}\psi'
(\frac{1}{\alpha_s} +\Delta)}=1+ \frac{b_1}{2\pi b_0} \alpha_s+
\frac{b_1}{2\pi b_0}\alpha^{2}_s\left[-\Delta+\frac12
+\frac{b_1}{2\pi b_0}\right]+$$
\be
+\frac{b_1}{2\pi b_0} \alpha^3_s\left[\Delta^2-\Delta+\frac16
+\left(\frac{b_1}{2\pi b_0}\right)
(1-2\Delta)+\left(\frac{b_1}{2\pi b_0}\right)^2\right]
\label{s.12} \ee

>From (\ref{s.12}) one can find the "theoretical" value of  $b_2$
and $b_3$, namely
\be
b^{(th)}_2= 16 \pi b_1\left(\frac12-\Delta+\frac{b_1}{2\pi
b_0}\right)= 3173.4-16\pi b_1\Delta\label{39} \ee $$ b^{(th)}_3=
64 \pi^2 b_1[\Delta^2-\Delta+\frac16 +2\gamma
\left(\frac12-\Delta\right) +\gamma^2],~~ \gamma= \frac{b_1}{2\pi
b_0}. $$
 Now since $\psi(z)$ has poles at
$z=0,-1,-2,...,$ $\psi\left(\frac{1}{\alpha_s}+\Delta\right)$ has
poles at the values of $\alpha_s$
\be
\alpha_s=-\frac{1}{\Delta},
-\frac{1}{1+\Delta},-\frac{1}{2+\Delta},... \label{40}\ee

Correspondingly $\psi'(z)$ has poles of second order, and as a
consequence of (\ref{s.1}) $\beta(\alpha_s)$ has zeros at the
values of $\alpha_s$ given in (\ref{40}).  These zeros condense to
$\alpha_s=0$ from the negative side.

Poles of $\beta (\alpha_s)$ are defined by the zeros of the
denominator in (\ref{s.1}), i.e. given by

\be
1=\frac{b_1}{2\pi b_0} \psi' \left(
\frac{1}{\alpha_s}+\Delta\right). \label{41}\ee

For $\Delta>0$ and $\alpha_s>0$ there is one root of (\ref{41}) at
$\alpha_s=\alpha_s^{(\Delta)}$, if $\Delta\leq \Delta_0$ where
$\Delta_0$ is defined by
\be
\psi'(\Delta_0)=\frac{2\pi b_0}{b_1}=1.355;
\Delta_0=1.145.\label{42}\ee

For $\Delta<\Delta_0$ one obtains one pole of $\beta(\alpha_s)$
given by (\ref{41}), while for $\Delta>\Delta_0$ there are no
poles of $\beta(\alpha_s)$ for $\alpha_s\geq 0$, and
$\beta(\alpha_s)$ is monotonically  decreasing function.

For $\alpha_s<0$ one has infinite number of poles of
$\beta(\alpha_s)$ which lie between zeros of $\beta(\alpha_s),$
described by Eq. (\ref{40}).

We now come to the analytic properties of $\alpha_s(Q^2)$, having
in mind parametrization (\ref{s.8}) for $\Pi(Q^2)$ in $ e^+e^-$
annihilation.

Poles of $\alpha_s$ are to be found from the equation
\be
\ln\frac{m^2}{\Lambda^2} +\psi\left( \frac{Q^2+M^2_0}{m^2}\right)
+ \frac{2b_1}{b_0^2} \psi \left( \frac{1}{\alpha_s}+\Delta\right)
=0 \label{43}\ee and we require that for  $Re Q^2\geq 0$ there
should be no poles of $\alpha_s(Q^2)$, which gives condition
\be
\psi\left(\frac{M^2_0}{m^2}\right) +\ln \frac{m^2}{\Lambda^2}
+\frac{2b_1}{b_0^2} min\psi\left( \frac{1}{\alpha_s}
+\Delta\right) \geq 0\label{44}\ee

Now $\psi(z)>0$ for $z>1.46$, and for  $m^2= M^2_0=1 GeV^2;
\Lambda=0.37 GeV$  (the accurate calculations in \cite{4,15,16}
strongly prefer $M_0=1$ GeV) one obtains for the sum of two first
terms in (\ref{43}) the  value \be
\ln\frac{m^2}{\Lambda^2}+\psi\left( \frac{M^2_0}{m^2}\right) =
1.99-0.577=1.41 \label{45}\ee

Finally one can see that for $\Delta>\Delta_0$ one has
$\psi(\Delta)>-0.5$ and the l.h.s. of (\ref{44}) is positive, and
hence for all $Q^2>0$ also l.h.s. of Eq.(\ref{43}) is always
positive. In this way we have proved that $\alpha_s(Q^2)$ has no
poles for all $Q^2>0$ if $\Delta>\Delta_0$.

The resulting form of $\alpha_s(Q^2)$ valid in all $Q^2$ plane is
now
\be
\alpha_B=\frac{4\pi}{b_0\left[\ln\frac{m^2}{\Lambda^2}
+\psi\left(\frac{Q^2+M^2_0}{m^2}\right)+\frac{2b_1}{b^2_0}
\psi\left( \frac{1}{\alpha_B}+\Delta\right)\right]}\label{46}\ee
where $\Delta>\Delta_0=1.145$. For $Q^2<0$ the function
$\alpha_s(Q^2)$ in (\ref{46}) has zero's and poles which not
necessarily coincide with poles of
$\psi\left(\frac{Q^2+M^2_0}{m^2}\right)$.

But this is as it should be, since the amplitude $\Pi(Q^2)$ must
have poles at the final positions
\be
-Q^2=(M^{(n)})^2,~~ n=0,1,2,...\label{47}\ee which take into
account valence (and Coulomb) gluon effects.

\section{A simple model for constructing perturbation series}

In this section we shall consider a simple model which is to
illustrate the analytic properties of separate terms of BPT and of
its infinite sum. Our discussion in this section will be formally
similar to the lectures \cite{20} where old-fashioned perturbation
theory was used for SPT, except that we use confined states
instead of free quarks and gluons.
 To this end we define a
quantum-mechanical Hamiltonian $H_0$ which produces bound states
(equivalent of infinite number of mesonic states) and two
second-quantized operators $V_1(t)$ and $V_2(t)$ which create
additional valence gluons, to be bound together with quark and
antiquark in a sequence of hybrid states. We define the total
Hamiltonian
\be
H=H_0+V_1(t)+V_2(t)\label{4.1}\ee and \be V_1(t)=e^{i\omega_1
t}v_1a^++h.c.\label{4.2}\ee

\be V_2(t)=e^{i\omega_2 t}v_2b^+a^+a+h.c.\label{4.3}\ee where
$a,a^+$ and $b,b^+$ are annihilation  (creation operators). We do
not specify the $d$-space dependence of $v_1(x), v_2(x); t$- is
the common time of the instant form of dynamics. It is clear that
$V_1, V_2$ are prototypes of the terms $\bar \psi\hat a \psi$ and
$a^2\partial a$ in the QCD Lagrangian, here to simplify matter we
disregard in $a_\mu(\vex)$ or $A_\mu(\vek)$ dependence on
polarization $\mu$ and momentum $\vek$, but introduced $b,a$ to
distinguish gluon emission by quarks (in $V_1$) and by gluons (in
$V_2)$.

We assume that $H_0$ has only discrete spectrum of mesons
(confinement), so for the Green's function we have
\be
i\frac{\partial}{\partial t} G_0(t)=H_0(t) G_0;~~
G_0(t)=\sum_n\varphi_n^{(0)}\varphi_n^{(0)+}e^{-iE_n^{(0)}t}\label{4.4}\ee
while for the Fourier transform one has
\be
G_0(E)=\int^\infty_0 e^{-iET}G_0(t)
dt=i\sum_n\frac{\varphi_n^{(0)}\varphi_n^{(0)+}}{E-E_n^{(0)}}.\label{4.5}\ee

For the full Green's function one can write
\be
i\frac{\partial}{\partial t} G(t)=(H_0+V(t))G(t),~~
V(t)=V_1(t)+V_2(t)\label{4.6}\ee with the solution \be G(q_1,
q_2;t)=<q_1|e^{-i\int^t_0(H_0+V(t')) dt'}|q_2>.\label{4.7}\ee

The perturbative series looks like $$ G(t)=G_0+\int G_0
(t-t_1)(-i)V(t_1) G_0(t_1) dt_1+$$ \be +\int\int G_0(t-t_1)(-i)
V(t_1) dt_1 G_0(t_1-t_2)(-i) V(t_2) G_0(t_2) dt_2+...
\label{4.8}\ee

For the energy-dependent Green's function $G(E)$  the perturbative
series can be written taking into account  the vacuum state for
$a,b$ operators \be a|vac> =b|vac>=0 \label{4.9} \ee which yields
$$ G(E)=G_0(E)+G_0(E)(-i\tilde V_1) G_0(E-\omega_1)(-i\tilde V_1)
G_0(E)$$
 $$+G_0(E)(-i\tilde V_1) G_0(E-\omega_1) (-i\tilde V_2)
G_0(E-\omega_1-\omega_2)(-i\tilde V_2) G_0(E-\omega_1)(-i\tilde
V_1)G_0(E)+ $$
 $$+G_0(E)(-i\tilde V_1) G_0(E-\omega_1) (-i\tilde V_1)
G_0(E)(-i\tilde V_1) G_0(E-\omega_1)(-i\tilde V_1)G_0(E)+ $$
\be
 +G_0(E)(-i\tilde V_1) G_0(E-\omega_1) (-i\tilde V_1)
G_0(E-2\omega_1)(-i\tilde V_1) G_0(E-\omega_1)(-i\tilde
V_1)G_0(E)+ ...\label{4.10}\ee

If one introduces the "gluon propagator"
\be
K_1(E)=(-i\tilde V_1) G_0(E-\omega_1) (-i\tilde
V_1)\label{4.11}\ee then the partial summation of diagrams with
gluons emitted by quarks only yields \be
G^{(K_1)}(E)=G_0(E)+G_0(E)K_1(E)G_0(E)+...=
G_0(E)\frac{1}{1-K_1G_0}.\label{4.12} \ee

Similarly for the "gluon-b" propagator
\be
K_2(E)=(-i\tilde V_2) G_0(E-\omega_1-\omega_2)(-i\tilde
V_2)\label{4.13}\ee and the "full gluon propagator" $$
G^{(K_2)}(E)=K_1(E)+(-i\tilde
V_1)G_0(E-\omega_1)[1+K_2G_0(E-\omega_1)+ $$
\be
(K_2G_0(E-\omega_1))^2+...] (-i\tilde V_1)= (-i\tilde V_1)
G_0(E-\omega_1) \frac{1}{1-K_2 G_0(E-\omega_1)}(-i\tilde
V_1)\label{4.14} \ee

 One should note that $\tilde V_i$ are operators and denominators
 in (\ref{4.12}) and (\ref{4.14}) are formal operator expressions. Now
 to analyze the perturbative expansion (\ref{4.10}) or its partial sums
 (\ref{4.12}),(\ref{4.14}) one should specify the unperturbed spectrum
 (\ref{4.5}) and define matrix elements  of $\tilde V_i$ between
 unperturbed eigenfunctions $\varphi_n^{(0)}$. We shall assume for
 simplicity that that the spectrum is linear in $n$,
 \be
 E_n^{(0)} = M_0+mn, n=0,1,...
 \label{62}
 \ee and for $V^{(i)}_{nk} \equiv (\varphi_n^{+(0)} \tilde
 V_i\varphi_k^{(0)})$ one can assume that at large $n,k$ this
 matrix element factorizes, $V_{nk}^{(i)}= c_n^{+(i)}c_k^{(i)}$.

 As a consequence one has
 \be
 G^{(K_1)} (E)=\sum_n\frac{\varphi_n c_n^{(1)+}}{E-E_n}
 \Omega_1(E-\omega_1)\frac{1}{(1-g(E)\Omega_1(E-\omega_1))}\sum_n\frac{c_{n'}^{(1)}\varphi_{n'}^+}
{E-E_{n'}} \label{63} \ee where we have defined
\be
\Omega_1(E-\omega_1)\equiv-\sum_k\frac{|c_k^{(1)}|^2}{E-E_k-\omega_1},\label{64}\ee
\be
g(E)\equiv \sum_n\frac{|c_n^{(1)}|^2}{E-E_n}\label{65} \ee

It is clear that $\Omega_1(E-\omega_1)$ has poles at the "gluon
excited" i.e. "hybrid state" masses $E=E_k+\omega_1$, however the
partial sum $G^{(K_1)}(E)$ has shifted poles due to vanishing of
denominator in (\ref{63}), in addition to unperturbed poles at
$E_n$.

A similar situation occurs for $G^{(K_2)}(E)$. If one defines
\be
\Omega_2(E-\omega_1)=\sum_k\frac{|c_k^{(2)}|^2}{E-E_k-\omega_1}\label{66}
\ee
\be
\Pi_2(E-\omega_1-\omega_2)=-\sum\frac{|c_k^{(2)}|^2}{E-E_k-\omega_1-\omega_2}\label{67}
\ee then one can rewrite (\ref{4.14}) as
\be
G^{(K_2)}(E)=\frac{(-i\tilde V_1)|\varphi_n>}{E-E_n-\omega_1}
\{\delta_{mn}+ c_n^{+(2)}\frac{\Pi_2(E-\omega_1-\omega_2)
c_m^{(2)}}{1-\Omega_2(E-\omega_1)\Pi_2(E-\omega_1-\omega_2)}\}\frac{<\varphi_m|(-i\tilde
V_1)}{E-E_m-\omega_1}.\label{68} \ee

This exercise was undertaken to illustrate that the "full gluon
propagator" $G^{(K_2)}(E)$ has in addition to unperturbed
"one-gluon poles" $\frac{1}{E-E_n-\omega_1}$ also shifted
"two-gluon poles" due to vanishing of denominator in (\ref{68}),
but not two-gluon poles" themselves.

To make our model more realistic and finally to incorporate in it
QCD features, one should make several steps. Firstly, one must
take into account negative energy states, corresponding to the
background in time motion, i.e. as in the Feynman propagator we
replace (cf.\cite{20})
\be
\frac{1}{E-E_n}\to \frac{1}{E^2-E^2_n} \to
\frac{1}{s-(M_n^{(0)})^2}.\label{69}\ee Correspondingly, the
hybrid propagators take the form
\be
\frac{1}{E-\omega_1-E_n}\to
\frac{1}{s-(M_n^{(1)})^2},~~\frac{1}{E-\omega_1-\omega_2-E_n}\to
\frac{1}{s-(M_n^{(2)})^2} \label{70} \ee where $M_n^{(1)},
M_n^{(2)}$ are hybrid masses with one and two gluons respectively.

Now one must take into account the structure of $\tilde V_1,
\tilde V_2$ operators in QCD. Both describe emission of one gluon
from quark and gluon line respectively, and both are local field
operators, whereas wave functions refer to the nonlocal
instantaneous bound state objects, namely, mesons $(q\bar q)$,
1-gluon hybrid $(q\bar q g)$ and 2-gluon hybrid ($q\bar q gg)$. We
shall denote the full set of the corresponding eigenfunctions as
$\varphi^{(0)}_n(\vex), \varphi^{(1)}_{n,\nu_1} (\vex, \vey)$ and
$\varphi^{(2)}_{n,\nu_1\nu_2} (\vex, \vey,\vez)$. Note that the
string pieces between quarks and gluons formed by background field
are straight lines, given by vectors $\vex,\vey$ in
$\varphi_n^{(1)}$, and by $\vex, \vey, \vez$ in $\varphi_n^{(2)}$,
and they contribute to the rotational and vibrational
eigenenergies. We also note that the string vibration in this
background picture is described by the full set of one-, two- and
many-gluon hybrid states.

Here we have introduced in hybrid eigenfunctions additional lower
indices $\nu_1$ and $\nu_1, \nu_2$ to describe the quantum numbers
of additional bound gluons, note that $\nu_i$ are multidimensional
vectors with integer components. Correspondingly hybrid masses
depend on those quantum numbers;
\be
M_n^{(0)}= M^{(0)} (n_r, L; J^{PC})\label{71}\ee
\be
M_n^{ (1)}=M^{(1)}(n_r, L; n'_r, L'; J^{PC})\label{72}\ee
\be
M_n^{ (2)}=M^{(2)}(n_r, L, n'_r, L', n_r^{\prime\prime},
L^{\prime\prime}; J^{PC}).\label{73}\ee

Here $n_rL$ refer to the radial quantum number and angular
momentum of quark with respect to antiquark,
 each additional gluon has its own $n_r, L$ with respect to
 neighboring quark or gluon.
  We have not taken into account yet spin degrees of freedom of
  $q,\bar q$ and gluons, which create fine and hyperfine structure
  of levels, and we shall neglect those, since we are interested
  in the dominant contributions to the masses at large
  excitations, and our final goal is the result of summation
  over all excitations in $G(E)$ and establishing of some relation
  between SPT and BPT, which requires analysis of high excited
  states in the sums.

  Now the local character of the operators $\tilde V_1, \hat V_2$
  mentioned above leads to the following properties of matrix
  elements (since $\tilde V_i$ acts only at one of the end points
  of vectors $\vex\vey$, it does not influence quantum wave
  function, unless $\tilde V_i$ contains derivative in coordinate
  as in $\tilde V_2$, and in this case it changes quantum numbers
  by a discrete number):
$$ <n_r,L |\tilde V^{(1)}| \bar n_r, \bar L; n'_r, L'>\sim
\delta_{n_r\bar n_r} \delta_{L\bar L}$$
\be
<n_r,L; n'_r, L' |\tilde V^{(2)}| \bar n_r, \bar L, \bar n'_r,
\bar L'; n_r^{\prime\prime}, L^{\prime\prime}>\sim \delta_{n_r\bar
n_r} \delta_{L\bar L}\delta_{\bar n'_r n'_r} \delta_{\bar L' L'}.
\label{74} \ee Moreover  we shall assume that $(M_n^{(i)})^2$
depends on all quantum numbers linearly,
\be
(M_n^{(i)})^2= m^2(a\sum_i n_r^{(i)} + b\sum L^{(i)})\label{75}
\ee Let us now apply these rules to the computation of the "full
gluon propagator" $G^K_2(E)$ and try to find a correspondence with
the standard perturbative calculations.

To this end we assume that the sum entering in $K_1(E)$ is
converging, i.e.
\be
<n|K_1(E)|m>= \sum_{n,\nu}
 \frac{<n|-i\tilde V_1| \bar n, \nu ><\bar n, \nu|-i \tilde V_1| m>}{Q^2
 +M^2(\bar n, \nu)}\sim
\frac{1}{Q^2} \label{76} \ee where we denote $n(\nu)=n_r, L(n'_r,
L')$ and replace $s\to - Q^2$.

The same type of expression for $K_2(E)$ yields
\be
<n,\nu|K_2(E)|\tilde n, \tilde \nu>=\sum_{\bar n, \bar \nu,\nu'}
\frac{<n,\nu|-i \tilde V_2| \bar n, \bar \nu, \nu'><\bar n, \bar
\nu, \nu'|-i\tilde V_2 |\tilde n, \tilde \nu>}{Q^2+ M^2(\bar n,
\bar \nu, \nu')}. \label{77} \ee

 Due to the
 properties (\ref{75}) the sum in (\ref{77}) is actually a double
 sum in, say, $L'_\nu, n'_r$, while $M^2$ depends on them
 linearly  and we assume at this point that coefficients in (\ref{77})
 do not depend on $\nu'$. Therefore one must regularize this
 expression, making one subtraction at $s=0$, and renormalize the
 retaining  logarithmic divergence. This is in full correspondence
 with the standard  perturbative calculation of  one gluon loop
 correction to gluon propagator. As a result $K_2(E)$ aquires the
 form
 \be
 <n,\nu|k_2(E)|\tilde n,\tilde \nu>\sim Q^2\ln\left
 (\frac{Q^2-M^2(n,\nu)-\delta M^2}{\mu^2}\right) \delta_{n\tilde n}
 \delta_{\nu\tilde \nu}\label{78}
 \ee
 where
 \be
 \delta M^2= M^2_{min} (n,\nu, \nu') -M^2(n,\nu)\label{79}\ee
 and minimization is taken with respect to $\nu'$.

 Now one can consider $G^K_2(E)$ extracting coupling constant $g_0$
 from $\tilde  V_1, \tilde V_2$ one has a series (at large $s$)
 \be
 G^{(K_2)}(s)\sim \frac{g^2_0}{Q^2} - C\frac{g^4_0}{Q^2} \ln \frac{Q^2+\bar
 M^2}{\mu^2}+...\sim \frac{1}{Q^2}
 \frac{g^2_0}{\left(1+g^2_0C\ln\frac{Q^2+\bar M^2}{\mu^2}\right)}.\label{80}\ee
Expression (\ref{80}) is similar to that of the  running coupling
constant, but with the mass $\bar M^2$, corresponding to the
freezing coupling constant derived in \cite{12} and  studied later
in \cite{13,14}. We note that logarithms in (\ref{80}) appear as a
result of summation over the string like spectrum (\ref{75}) with
the coefficients asymptotically constant, which is the property of
matrix elements of eigenfunctions of linear interaction \cite{14}.
Therefore this property and the correspondence found above can be
only an asymptotics, while at finite $s$ there are corrections,
which should be taken into account. In the following chapter we
look into this problem more carefully.

\section{
Nonasymptotic terms in the spectral sums}

We start with the definition of the photon vacuum polarization
function $\Pi(q^2)$ and after a short review of asymptotic
correspondence between the spectral sum and perturbative
expansion, we shall look more carefully into the correction terms
and comparison between OPE and spectral sum.

We define the photon vacuum polarization function $\Pi(q^2)$ as a
correlator of electromagnetic currents for the process $e^+e^-\to
$ hadrons in usual way
\be
-i\int d^4xe^{iqx}<0|T(j_{\mu}(x)j_{\nu}(0)|0>= (q_{\mu }
q_{\nu}-g_{\mu\nu}q^2)\Pi(q^2),\label{81} \ee where the imaginary
part of $\Pi$ is related to the total hadronic ratio $R$ as
\be
R(q^2)=\frac{\sigma(e^+e^-\to
hadrons)}{\sigma(e^+e^-\to\mu^+\mu^-)}= 12\pi e Im
\Pi(\frac{q^2}{\mu^2},\alpha_s(\mu))\label{82} \ee

 There are two standard approaches to calculate
 $\Pi(q^2)$;
 first, when purely perturbative expansion is used,
 which is now known to the order $0(\alpha^3_s)$ [21]. The
 second one is the OPE approach [6], which includes the
 NP contributions in the form of local condensates. For
 two light quarks of equal masses $(m_u=m_d=m)$ it
 yields for $\Pi(Q^2)$
 \be
\frac{1}{e^2_q}\Pi(Q^2)=-\frac{1}{4\pi^2}(1+\frac{\alpha_s}{\pi})ln
\frac{Q^2}{\mu^2}+\frac{6m^2}{Q^2}+\frac{2m<q \bar q>}{Q^4}
+\frac{\alpha_s<FF>}{12\pi Q^4}+...\label{83} \ee

To make explicit the renormalization of $\alpha_B$, one can write
   the perturbative expansion of the function $\Pi(Q^2)$ (23) as
   \be
   \Pi(Q^2)=\Pi^{(0)}(Q^2)+\alpha_B\Pi^{(1)}(Q^2)+\alpha_B^2\Pi^{(2)}
   (Q^2)+...
\label{84}   \ee

   We now again use the large $N_c$ approximation, in which case
   $\Pi^{(0)}$ contains only simple poles in $Q^2$ [27,28]:
   \be
   \Pi^{(0)}(Q^2)=\frac{1}{12 \pi^2}
   \sum^{\infty}_{n=0}\frac{C_n}{Q^2+M^2_n}\label{85}
   \ee
   and the mass $M_n$ is an eigenvalue of the Hamiltonian $H^{(0)}$,
   which contains only quarks and background field $B_{\mu}$,
   \be
   H^{(0)}\Psi_n=M_n\Psi_n,\label{86}
   \ee

      In what follows we are mostly interested in the long--distance
   effective Hamiltonian which obtains from $G_{q\bar q}$ for large
   distances, $r\gg T_g$, where $T_g$ is the gluonic correlation
   length of the vacuum, $T_g\approx 0.2 fm$, [19]:
   \be
   G_{q\bar q}(x,0)=\int
   DB\eta(B)tr(\gamma_{\mu}G_q(x,0)\gamma_{\mu}G_q(0,x))=
   <x|e^{-H^{(0)}|x|}|0>\label{87}
   \ee
    At these distances one can neglect in $G_q$ the  quark spin
    insertions $\sigma_{\mu\nu}F_{\mu\nu}$ and use the area law:
    \be
    <W_C>\to exp (-\sigma S_{min})\label{88}
    \ee
    where $S_{min}$ is the minimal area inside the loop $C$.

    Then the Hamiltonian in (\ref{87}) is readily obtained by the method of
    [22]. In the c.m. system for the orbital momentum $l=0$ it has the
    familiar form:
    \be
    H^{(0)}=2\sqrt{\vec p^2+m^2}+\sigma r + const,\label{89}
    \ee
    where constant appears due to the  self-energy parts of quarks  [23],
    while for $l=2$ a small correction from the rotating string
    appears [22], which we neglect in the first approximation.

    Now we can use the results of the quasiclassical analysis of
    $H^{(0)}$ [24] where the values of $M_n,C_n$ have been  already
    found.

    These results can be represented as follows
    $(n=n_r+l/2,n_r=0,1,2,...,l=0,2)$,
    \be
    M^2_n=2\pi\sigma (2n_r+l)+M_0^2,\label{90}
    \ee
    where $M_0^2$ is a weak function of  quantum numbers $n_r,l$
    separately,
    comprising  the constant term of (\ref{89}); in what follows we shall
    put it equal to the $\rho$--meson mass, $M_0^2\simeq m^2_{\rho}$ (see [23] for more discussion).

     For $C_n$ one obtains quasiclassically [24]
     $$
     C_n(l=0)=\frac{2}{3}e^2_qN_cm^2,~~m^2\equiv 4\pi\sigma,
     $$
     \be
     C_n(l=2)=\frac{1}{3}e^2_qN_cm^2,\label{91}
     \ee
     Using asymptotic expressions (\ref{90}),(\ref{91}) for $M_n,C_n$ and starting
     with $n=n_0$, one can write
     \be
     \Pi^{(0)}(Q^2)=+\frac{1}{12\pi^2}\sum^{n_0-1}_{n=0}
     \frac{C_n}{M^2_n+Q^2} -
     \frac{e^2_qN_c}{12\pi^2}\psi(\frac{Q^2+M_0^2+n_0m^2}{m^2})\label{92}
     \ee
     $$
     + ~divergent~~ constant,
     $$
     where we have used the equality
     \be
     \sum^{\infty}_{n=n_0}\frac{1}{M^2_n+Q^2}=
     -\frac{1}{m^2}
     \psi(\frac{Q^2+M_0^2+n_0m^2}{m^2})\label{93}
     \ee
     $$
     + ~divergent~~ constant
     $$
     and
     $\psi(z)=\frac{\Gamma'(z)}{\Gamma(z)}$.

     In (\ref{92}) we have separated the first $n_0$ terms to
     treat them nonquasiclassically, while keeping for
     other states with $n\geq n_0$ the quasiclassical
     expressions (\ref{90}),(\ref{91}). In what follows, however, we
     shall put $n_0=1$ for simplicity. It was shown in [14]
      that even in this case our results
     reproduce $e^+e^-$ experimental data with good
     accuracy.

Consider now the asymptotics of $\Pi^{(0)}(Q^2)$ at large $Q^2$.
     Using the asymptotics of $\psi(z)$:
\be
\psi(z)_{z\to\infty}=ln
       ~z-\frac{1}{2z}-\sum^{\infty}_{k=1}\frac{B_{2k}}{2kz^{2k}}\label{94}
       \ee
       where  $B_n$ are Bernoulli numbers, from (\ref{92}) one obtains
       \be
\Pi^{(0)}(Q^2)=-\frac{e^2_q N_c}{12\pi^2}ln
\frac{Q^2+m^2}{\mu^2}+0(\frac{m^2}{Q^2})\label{95}
       \ee

One can easily see that this term coincides at $Q^2\gg M_0^2$ with
the first term in OPE (22) -- the logarithmic one. Taking the
imaginary part of (39) at $Q^2\to -s$ one finds
\be
R(q^2)=12\pi Im\Pi^{(0)}(-s)=N_ce^2_q, \label{96}\ee i.e. it means
that we have got from $\Pi^{(0)}$ the same result as for the free
quarks. This fact is the explicit manifestation of the
quark--hadron duality.

We can now identify $\Pi^{(0)}(Q^2)$ as
\be
\Pi^{(0)}(Q^2)= -\frac{e^2_qN_c}{12\pi^2}
\psi\left(\frac{Q^2+M^2_0}{m^2}\right)\label{97} \ee and compare
this expression with the OPE series  (\ref{83}).

To this end one can use the asymptotic expansion of $\psi(x)$ $$
\psi(z)_{z\to\infty} = \ln z -\frac{1}{2z}-\sum^\infty_{k=1}
\frac{B_{2k}}{2kz^{2k}}=$$
\be
=\ln z-\frac{1}{2z} -\frac{1}{12 z^2} +\frac{1}{120 z^4}
-\frac{1}{252 z^6}+...\label{98} \ee and the expansion of
\be
\psi(z)=-C-\sum^\infty_{k=0} \left(
\frac{1}{z+k}-\frac{1}{k+1}\right),~~ C=0.577.\label{99} \ee

Two features are immediately seen when comparing expansion
(\ref{98}) with the OPE (\ref{83}) (see also discussion in
\cite{13, 14}) i) for $m=0$ the term $O(1/Q^2)$ is absent, since
one cannot construct a local gauge-invariant operator of dimension
mass squared.   One should stress, that this type of operator may
appear from the interference of perturbative and nonperturbative
contributions \cite{25} and is welcome on phenomenological grounds
\cite{26}. However in the framework of BPT accepted in this paper
the  term $\Pi^{(0)} (Q^2)$ is solely due to NP contributions, and
cannot have OPE terms $O(1/Q^2)$.

ii) The generic terms in $\Pi^{(0)}(Q^2)$ from expansion
(\ref{97}) have the magnitude $\frac{m^{2n}}{Q^{2n}}$, while the
corresponding OPE terms are much smaller, e.g. one should compare
$-\frac{\pi}{3} \frac{\alpha_s\lan FF\ran}{Q^4} \approx
-\frac{0.04{\rm~ GeV}^4}{Q^4}$ (OPE) with
$-\frac{m^4}{12Q^4}\approx -\frac{0.5 {\rm~GeV}^4}{Q^4}$ (spectral
decomposition (\ref{97})).

In the remaining part of this section we shall discuss possible
solutions of problems  i) and ii).

Concerning the point i), one should say that (\ref{97}) is an
approximate expression for $\Pi^{(0)}$ valid for large $Q^2$, i.e.
reproducing correctly the logarithmic term in (\ref{95}). However,
if one interested in the next asymptotic terms, one should take
into account corrections to $C_n$ and $ M_n$ in (\ref{92}).

The first major correction comes from the fact that we have taken
into account in (\ref{93}) that levels with $n_r+1, l=0$ and $n_r,
l=2$ are degenerate (in the lowest approximation when one neglects
spin-dependent force) and hence used the sum of coefficients
(\ref{91}) for $C_n$ in (\ref{92}). However it is not true for the
$\rho$-meson, where coefficient with $l=0$ yields $\frac23$ of
$C_n$. Therefore one should use instead of $\psi(z)$ in
(\ref{93}), (\ref{97}) the corrected expression
\be
\psi(z)\to \tilde \psi(z)\equiv \psi (z)
+\frac{1-\gamma}{z}\label{100}\ee where $\frac{1-\gamma}{z}$
cancels the corresponding pole (with $k=0$) in (\ref{99}) and
replaces it with the correct  coefficient
$\gamma=\gamma_0=\frac23$.

Moreover, in the analysis \cite{14} it was realized that the value
$\gamma_0=\frac23$ (which does not take into account radiative
correction to the $\rho$ width) yields some 10\% discrepancy in
the  leptonic width of $\rho$ meson. Now the destiny of the
$\frac{1}{q^2}$ term in the asymptotic expansion (\ref{98})
depends on the exact value of $\gamma$. Indeed, using (\ref{98}),
one obtains the expansion
\be
\tilde \psi (z)=\ln z+\frac{1}{z} (1-\gamma-\frac12)
+O(\frac{1}{z^2})\label{101}\ee where $z=\frac{Q^2+M^2_0}{m^2} \to
\frac{ Q^2}{m^2}$.

One can see that the requirement of the absence of $O(1/Q^2)$
terms implies the following condition of $\gamma$:
\be
\gamma=\frac23 \kappa_{rad} =\frac12\label{102}\ee where
$\kappa_{rad}=1-\Delta \kappa$, and $\Delta\kappa$ is the
radiative correction due to gluon exchanges in the leptonic width
of $\rho$-meson. It is clear that condition is in the correct
ballpark and one should take into account other possible
corrections, to be discussed below.

Till now we have considered the lowest approximation for
 for $M^2_n$ and $C_n$, Eqs.(\ref{90}), (\ref{91}),  where $M^2_n$
 is linear in $n$ and $C-n$ is $n$ -independent. However, there
 are corrections to this quasiclassical behaviour \cite{27},
 yielding
 \be
 M^2_n= 4\pi \sigma n + M^2_0+\frac{b}{n}
 \label{103}\ee
 $$
 C_n= C_n^{(0)} \left( 1+\frac{a}{n}\right)
 $$
 where $C_n^{(0)}$ is given in (\ref{91}). To see the effect of
 these corrections we first put $b=0$ and consider the sum (cf.
 Eq.(\ref{93})), which can be  expressed through $\psi(z)$ again

 \be
 \sum^\infty_{n=n_0=1}
 \frac{(1+\frac{a}{n})}{M^2_0+m^2+ Q^2}
 = \label{104}\ee
 $$-\left\{ \frac{1}{m^2}\left(
 1-\frac{am^2}{M^2_0+Q^2}\right)\psi
 \frac{(M^2_0+Q^2+m^2)}{m^2}-
 -C\frac{a}{M^2_0+Q^2}\right\},
 C=0.577.
$$ Thus we see that the correction $\frac{a}{n}$ yields a power
correction $\frac{am^2}{M^2_0+Q^2}$ which modifies asymptotic
expansion of $\psi(z)$ (\ref{98}) and can occasionally diminish
some term to bring them closer in agreement with in the OPE
(\ref{83}). The same is true also for the mass correction $b/n$ in
(\ref{103}), which yields power correction
$\frac{bm^2}{(M^2_0+Q^2)^2}$ as a factor in expression like
(\ref{104}). However since coefficients in (\ref{98}) are fast
diminishing, the overall cancellation in all terms seems to be
unprobable and therefore the problem ii) remains unsolved.

\section{Analytic properties of $\alpha_B(s)$}

In this section we shall discuss the analytic structure of the BPT
in the limit $N_c\to \infty$ in the whole complex plane of
$-s=Q^2$ and establish connection between our results and SPT. In
what follows the basic role will be played by the function
$\Pi(Q^2)$ and we shall use this example as a typical one,
generalizing to other amplitudes at the end of this section.

Our starting point is the BPT for $\Pi(Q^2)$, Eq.(\ref{84}).
Analytic properties of  $\Pi^{(0)}(Q^2)$ are given in
Eq.(\ref{85}), showing a sequence of purely nonperturbative poles
\be
s=-Q^2=(M^{(0)}_n)^2, n=0,1,2,... \label{105}\ee

Inclusion of $\alpha_B$ to all orders makes the following changes
with the unperturbed spectrum (\ref{105}). The gluon exchanges
play two different roles: a) instantaneous Coulomb interaction
shifts NP poles b) gluons $a_\mu$ propagating in the confining
film form hybrid states. Correspondingly the NP poles are mixed
and shifted by the hybrid states.  Therefore one has the following
hierarchy of the shifts of original NP meson masses
\be
M_n^{(0)}\to M_n^{(0)} (O((\alpha_B)^K))\equiv
M_n^{(0)}(k)\label{106}\ee The one-gluon hybrid state (apart from
Coulomb shift, which we disregard for simplicity at this moment)
with mass $M^{(1)}_n$ is also shifted due to mixing with meson and
higher hybrid states, the mixing  being characterized by the power
of $\alpha_B$, and therefore one has
\be
M_n^{(1)} (O((\alpha_B)^k))\equiv M_n^{(1)} (k) \label{107}\ee In
a similar way one has shifted states $M_n^{(m)}(k)$ for m-hybrid
configurations (with $m$ gluons).

As it is clear from the schematic example of section 4, the terms
$O(g^2_0)$ (the second term on the r.h.s.of (\ref{4.10})) refer to
1-gluon hybrid states and do not contain any analytic structure of
$\alpha_B(Q^2)$, since they are not yet renormalized (and of
course not yet RG improved, i.e.resummed).

This renormalization process, leading to the nontrivial dependence
of $\alpha_B(Q^2)$, starts with the diagrams of the order
$O(g^4_0)$ in SPT (and of the order $O(\tilde V_1^2\tilde V^2_2)$
 in the example of section 4 - see the third term on the r.h.s. of
 (57)).

 This term and its QCD analog has singularities of the double
 hybrid (the term $G_0(E-\omega_1-\omega_2)$ in (57)),
 corresponding to the gluon-loop diagram and accompanying diagrams
 in $ O(g^4_0)$ in SPT.

Next one can do a RG improvement of this $O(g^4_0)$ result, which
leads to the resummation presented in $G^K_2(E)$, Eq. (68). In
this approximation one can write
\be
\alpha_B=\frac{4\pi}{b_0[\ln \frac{m^2}{\Lambda^2} + \tilde \psi
(Q^2)]}\label{108}\ee where $\tilde \psi (Q^2)$ is a meromorphic
function with poles at the double-hybrid position, i.e. at
\be
s=M^{(2)}_n (k=0)\label{109} \ee In the particular case when
$\tilde \psi (Q^2)=\psi \left(\frac{Q^2+M^2_0}{m^2}\right)$ and
$M_0\equiv M_0^{(2)}(k=0)$ the  form (\ref{108}) coincides with
the one, suggested in \cite{14}.

In the large $Q^2$ limit one has
\be
\tilde \psi (Q^2) \to \ln \frac{Q^2+M^2_0}{m^2} \label{109a}\ee
and $\alpha_B$ has the same freezing form, which was extensively
used before in \cite{12}-\cite{16}:
\be
\bar\alpha_B=\frac{4\pi}{b_0\ln
\frac{Q^2+M^2_0}{\Lambda^2}}\label{110} \ee Now we can compare the
form (\ref{108}) with our model solution (\ref{46}).
 Indeed when $\frac{Q^2+M^2_0}{m^2}$ is large, one can keep in the
 denominator of (\ref{46}) the first two terms, and one has the
 form  coinciding with (\ref{108}). In the next approximation one
 can keep in (\ref{108}) the term $\frac{2b_1}{b_0^2}\psi \left(
 \frac{1}{\alpha_B}+\Delta\right)
$ which yields next order approximation, when $M_n^{(2)}(0)$ is
shifted to the higher order position $M_n^{(2)}(2)$ and so on.

Thus one can see that while the $O(g^4_0)$ term contains the
double hybrid poles, the RG improvement leads to the shifted
poles,  which do not correspond to the final physical poles, the
latter are the result of infinite resurmmation to all orders.

Let us now discuss the imaginary part of $\Pi(Q^2)$. It is clear
that $\Pi^{(0)}(Q^2)$ contains the NP poles. The term $O(g^2_0)$
(equivalent of the second term on the r.h.s. of (\ref{4.10}))
contains double NP poles from $ (G_0(E))^2$ and 1-gluon hybrid
poles from $G_0(E-\omega_1)$. The $O(g^4_0)$ term contains double
NP poles, double 1-gluon hybrid poles and 2-gluon hybrid poles -
the latter are associated with $\alpha_B$ in the lowest order. The
same type of classification goes on for higher terms. The outcome
for the expansion (\ref{84}) is that in the term $\alpha_B^{k}\Pi
^{(k)} (Q^2)$ part of singularities is associated with
$\alpha^k_B$ and another  part with $\Pi^{(k)}(Q^2)$.

Thus we see that the requirements of i) logarithmic behaviour of
$\tilde \psi (Q^2)$ at large $Q^2$ and ii) meromorphic analytic
properties in the whole $Q^2$ plane, leads to the one possible
representation of $\tilde \psi (Q^2)$;
\be
\tilde \psi (Q^2)=\psi\left(\frac{Q^2+M^2_0}{m^2}\right)
+\sum^N_{k=0} \frac{a_k}{Q^2+m^2_k}\label{111}\ee where $N$ is
finite, so that logarithmic asymptotics of $\psi(z)$ is not
modified.

Let us now compare our Eq.(\ref{s.6}) with the SPT, more
explicitly with Gell-Mann-Low equation, written in the form
adopted by Radyushkin \cite{9}.

\be
L=\ln\frac{Q^2}{\Lambda^2}
=\frac{4\pi}{b_0\alpha_s}+\frac{2b_1}{b_0^2} \ln
\frac{\alpha_s}{4\pi}+\tilde
\Delta+\frac{b_2b_0-8b_1^2}{2b_0^3}\frac{\alpha_s}{4\pi} +
O\left(\left(\frac{\alpha_s}{4\pi}\right)^2\right).\label{112}\ee
Here $\tilde \Delta$ is  a parameter of integration which fixed
the definition of $\Lambda$. This should be compared with our
Eq.(\ref{s.6}) which we can write in a similar form:
\be
\ln\frac{m^2}{\Lambda^2} + \psi\left(\frac{Q^2+M^2_0}{m^2}\right)=
\frac{4\pi}{b_0\alpha_s}
-\frac{2b_1}{b_0^2}\varphi\left(\frac{1}{\alpha_s}\right)
\label{113}\ee

It is clear that our  analysis with the expansion (\ref{s.7}) is
equivalent to the analysis of (\ref{112}) for large $Q^2$ and
small $\alpha_s$. As we have found in (\ref{39}) the predicted
values of $b_2^{(th)}, b_3^{(th)}$ depend on the value of $\Delta$
and keeping the choice
$\varphi\left(\frac{1}{\alpha_s}\right)=\psi\left(\frac{1}{\alpha_s}+\Delta\right)$,
one has from (\ref{39}) for $\Delta\geq \Delta_0=1.145$
\be
b_2^{(th)}<238.15:~~~b_3^{(th)}\geq 18266\label{114} \ee

To repair obtained values and get agreement with  $ \overline{MS}$
calculated values \cite{21} one can use the modified $\psi(z)$
function like that in (\ref{111}), namely
\be
\varphi\left(\frac{1}{\alpha_s}\right) =\psi \left(
\frac{1}{\alpha_s}+\Delta\right)+\sum^N_{k=0}\tilde
A_k\frac{\alpha_s}{1+\alpha_s(\Delta +k)}.\label{115}\ee It is
important that the position of  poles of $\psi(k)$ are  not
changed, the only change is in the coefficients of those pole
terms, which can change the expansion (\ref{s.11}), yielding
agreement with $\overline{MS}$ coefficients of expansion
(\ref{s.2}).

We consider now the  problem of analytic continuation of
$\alpha_s(Q^2)$ and $\alpha_B(Q^2)$ from the Euclidean region
$Q^2\geq 0$ to the time-like region $Q^2<0$.

One can see that the freezing form $ \overline{\alpha}_B$
(\ref{110}) is well defined for $Q^2\geq -M^2_0$, and has a
logarithmic branch point at $Q^2=-M^2_0$.

 However this is an
artefact of representation (\ref{111}), which is valid in the
asymptotics when $\frac{Q^2+M^2_0}{\Lambda^2}\gg 1$, otherwise one
should use the original form (\ref{108}).

Choosing it in the form
\be
\alpha_B=\frac{4\pi}{b_0[\ln\frac{m^2}{\Lambda^2}+\psi \left (
\frac{Q^2+M^2_0}{m^2}\right)]}\label{117}\ee one  can see that
$\alpha_B(Q^2)$ is defined in all complex plane $Q^2$ and has only
isolated poles there at zeros of the denominator in (\ref{117}).
These poles are the result of partial resummation of two-gluon
hybrid poles present in $\psi\left(\frac{Q^2+M^2_0}{m^2}\right)$
and lying at $s\equiv -Q^2=M^2_0+n m^2, n=0,1,2,...$ We note that
this resummation is the standard RG improvement of perturbative
series, where large logarithms  are summed up in the geometric
series (or else a result of solution of RG equations). The crucial
point
 is that most important physical thresholds are
contained in $\Pi^{(0)}(Q^2)$ and have nothing to do with
singularities of $\alpha_B(Q^2)$ (or of $\alpha_s(Q^2))$.

In a similar way the singularities which appear from the
asymptotic solution of (\ref{46}) for small  $\alpha_B$, namely
term $\ln \alpha_B(Q^2)$ does not produce additional logarithmic
cuts and is  a result of asymptotic expansion of Eq.(\ref{46}),
where in exact solution only meromorphic function $\alpha_B(s)$ is
obtained.

Thus as we discussed above the equation (\ref{46}) or its
generalized form
\be
\alpha_B=\frac{4\pi}{b_0\left [\ln\frac{m^2}{\Lambda^2} +\tilde
\psi(Q^2)+\varphi\left(\frac{1}{\alpha_B}\right)\right]}\label{118}\ee
where $\tilde \psi(Q^2)$ is given in (\ref{111}) and $
\varphi\left(\frac{1}{\alpha_B}\right)$ is given in (\ref{115}),
produce $\alpha_B(Q^2)$ which is meromorphic function of $Q^2$ in
the whole complex plane of $Q^2$.

Let us now look more carefully at the creation of imaginary part
of $\Pi(Q^2)$, which leads to the positive hadronic ratio $R(s)$.
We start with   $\Pi^{(0)}(Q^2)$ (\ref{85}) and (\ref{83}):
$\Pi^{(0)}$ in SPT and BPT coincide asymptotically, namely $$
\Pi^{(0)}(SPT)=-\frac{e^2_qN_c}{12\pi^2} \ln \frac{Q^2}{\mu^2}$$
\be
 \Pi^{(0)}(BPT)=-\frac{e^2_qN_c}{12\pi^2} \ln
 \frac{Q^2+M^2_0}{\mu^2}\label{119}
 \ee
Defining $s=-Q^2,~ arg (M^2_0-s)=0$ for $s<M^2_0$, one has for
large $s$
\be
Im\Pi^{(0)}_{BPT}=\frac{\Pi^{(0)}(s+i\varepsilon)-\Pi^{(0)}(s-i\varepsilon)}{2i}
= \frac{e^2_qN_c}{12\pi} \Theta (s-M^2_0)\label{120}\ee and for
SPT one has on the r.h.s. $\Theta(s)$ instead of
$\Theta(s-M^2_0)$.

Note however that the form (119) for $\Pi^{(0)}$ (BPT) is the
asymptotics at large positive $Q^2$ and in getting the result
(\ref{120}) the analytic continuation of the logarithmic
asymptotics was done unlawfully strictly speaking.

To calculate in  BPT more rigorously, one should use (\ref{85}) in
the region $s>M^2_0$, and one gets $$ Im\Pi^{(0)}(s+i\varepsilon)
= -\frac{e^2_qN_c Im}{12\pi^2} \psi\left(
\frac{-s+M^2_0}{m^2}\right) = +\frac{e^2_qN_c}{12\pi^2} Im
\sum^\infty_{k=0}\frac{m^2}{M^2_0-s+km^2}= $$
\be
 \frac{e^2_qN_c}{12\pi}
m^2\sum^\infty_{k=0}\delta (M^2_0+km^2-s) \label{121}\ee One can
introduce the average of $Im\Pi^{(0)}=(s+i\varepsilon)$ over some
energy interval (this is the "duality interval" discussed e.g. in
\cite{7}) comprising $N$ poles \be \lan
Im\Pi^{(0)}(s_0+i\varepsilon)\ran_N =\int^{s_0+Nm^2}_{s_0}\frac{ds
Im\Pi^{(0)}(s+i\varepsilon)}{Nm^2}=\frac{e^2_qN_c}{12\pi}\label{122}\ee
One can see that (\ref{122}) coincides with (\ref{120}) thus
justifying the procedure of direct analytic continuation of the
logarithms in (\ref{119}), when the averaging over the duality
interval is done.

One can also see that the standard SPT expression (\ref{119})
yields the correct answer for the same reason: the imaginary part
of asymptotics (both of SPT and BPT) coincides with the correct
averaged imaginary part. We note that for SPT this check is
impossible, since  the correct procedure is not available.

We now turn to the more delicate point
 of
analytic  continuation of $\alpha(s)$. In BPT, as was discussed
above, $\alpha_B(s)$is given by (\ref{118}) or in "one-loop
approximation" by (\ref{117}) and is a meromorphic function, hence
its analytic continuation is a direct and unique procedure. E.g.
for $\alpha_B$ (\ref{117}) separating one pole in $\psi\left(
\frac{Q^2+M^2_0}{m^2}\right)$, $M^2_p=M^2_0+nm^2$, one has in its
vicinity
\be
\alpha_B(s\approx M^2_{n}+i\varepsilon)
=\frac{4\pi}{b_0\psi'(z_0)(z-z_0)} \label{123}\ee where $z_0$ is
to be found from the equation
$\ln\frac{m^2}{\Lambda^2}+\psi(z_0)=0,$ and $z=
\frac{M^2_0-s}{m^2}.$ (If one uses exact model expression
(\ref{46}) instead of "one-loop approximation" Eq. (\ref{117}) the
r.h.s. of (\ref{123}) and subsequent (Eqs. (\ref{124}) and
(\ref{126}) should be multiplied by the constant factor
$(1-\frac{b_0}{4\pi} \psi' (\Delta))$, which effectively decreases
$Im\alpha_B$).
 Using relations from Appendix, one gets
\be
Im\alpha_B(s\cong M^2_n+i\varepsilon) =\frac{4\pi^2}{b_0}\frac{m^2
\delta(M^2_n-s-\delta_0m^2)}{\left[
\frac{1}{\delta^2_0(n)}+S_2(n)+\zeta(2)\right]}\label{124}\ee
where $\delta_0$ and $S_k(n)$ are defined in Appendix, $$
\delta^{-1}_0(n)=\ln\frac{m^2}{\Lambda^2}-C+S_1(n)$$
 Let us
now compare this exact procedure with the suggestive one, using
the asymptotic form of (\ref{117}), Eq.(\ref{110}). As before, we
have
\be
Im(\bar \alpha_B)\equiv \frac{\bar \alpha_B(s+i\varepsilon)-\bar
\alpha_B(s-i\varepsilon)}{2i}
=\frac{4\pi^2\Theta(s-M^2_0)}{b_0\{\ln^2|\frac{M^2_0-s}{\Lambda^2}|+\pi^2\}}\label{125}\ee

One can notice  on the other hand, that averaging over some
duality interval in (\ref{124}) yields
\be
\lan
Im\alpha_B\ran_{Nm^2}\cong\frac{4\pi^2}{b_0}\frac{1}{N}\sum^N_{k=0}
\frac{1}{\delta^{-2}(n+k)+\frac{\pi^2}{3}},\label{126}\ee and this
result as shown in Appendix asymptotically agrees with
(\ref{125}), specifically at large $s$ when the denominator in
(\ref{125}) grows. Now we turn to another definition of
discontinuity od $\alpha_B$ which is widely used in $e^+e^-$
annihilation and $\tau$ - decay. To make contact with SPT
calculations consider instead of $\Pi(Q^2)$ the so called Adler
function $D(Q^2)=\frac{Q^2d\Pi(Q^2)}{dQ^2}$ with the expansion
\be
D(Q^2)=\sum_qe^2_q\{ 1 +\frac{\alpha_s(Q^2)}{\pi}+ d_2\left(
\frac{\alpha_s(Q^2)}{\pi}\right)^2+d_3\left(
\frac{\alpha_s(Q^2)}{\pi}\right)^3+...\} \label{127}\ee where
coefficient $d_3$ was found in \cite{21}, and $d_2$ in \cite{28}

\be
d_2(\overline{MS}) = 1.986-0.115 n_f,~~ \bar d_3(\overline{MS})
=18.244-4.216 n_f+0.086 n^2_f\label{128}\ee

The procedure suggested in \cite{9} to go from $D(Q^2)$ to $R(s)$
is straightforward, provided one knows discontinuity of $\Pi(Q^2)$
(see discussion in \cite{9}, \cite{8} and recent publication
\cite{29})
\be
R(s)=\frac{1}{2\pi i} \int^{-s+i\varepsilon}_{-s-i\varepsilon}
D(\sigma) \frac{d\sigma}{\sigma}\label{129} \ee where integration
goes on both sides of the cut on real axis starting at  $s=0$.

To proceed one inserts in (\ref{129}) one - or two-loop expression
for $\alpha_s(Q^2)$ and as the result one obtains an expansion for
$R(s)$ \cite{9}
\be
R(s)=\sum e^2_q\left\{1+\sum_{k=1} d_k \Phi\left\{\left(\frac
{\alpha_s}{\pi}\right)^k\right\}\right\} \label{130}\ee where
$\Phi\{\}$ is the transformation defined by  (\ref{129}). In this
procedure the $\ln Q^2$ in  the denominator of $\alpha_s(Q^2)$
transforms into $\ln s$ with coefficients less than unity and
asymptotically tending to one.

Let us now consider what happens with our expression for
$\alpha_s(Q^2)$ (\ref{117}), which is valid not only for large
$Q^2$ but in the whole  $Q^2$ plane. We assume that in BPT the
function $D_B(Q^2)$ is given by the same expansion (\ref{127}),
where only $\alpha_s$ should be replaced by $\alpha_B(Q^2)$ and we
use (\ref{117}) for $\alpha_B(Q^2)$ to insert into (\ref{129}). In
this way one obtains, keeping the notation of \cite{9} \be
\Phi\left \{\left(\frac{
\alpha_B}{\pi}\right)^k\right\}=\frac{1}{2\pi
i}\left(\frac{4}{b_0}\right)^k\sum_{n=0}
\int^{s-i\varepsilon}_{s+i\varepsilon}\frac{
ds'}{[\psi'(z_0)(z-z_0)]^k s'}.\label{131}\ee

As shown Appendix, $\psi'(z_0)\approx
(\ln\frac{M^2_0+nm^2}{\Lambda^2})^2$. Here  the pole of the $k$-th
order is at $s'=M^2_0+nm^2-\delta_0(n)m^2$, and $n$ satisfies
condition $M^2_0+nm^2\leq s$.
 For arbitrary $k$ one obtains
\be
\Phi\left\{\left(\frac{\alpha_B}{\pi}\right)^k\right\} =0
\label{132}\ee This is a property of the integral in (\ref{131})
with the weight $\frac{ds}{s}$ that any function representable as
a finite sum of poles yields zero since
$\int^{s+i\varepsilon}_{s-i\varepsilon}\frac{dz}{z(z-s_n)^k}=0$.
However in the asymptotic region of $s=\bar s(1+i\gamma),~~\bar
s\gg m^2$,when number of poles is large, one can use asymptotic
form of $\alpha_B$,(\ref{111}), and the result is
\be
\Phi\left\{\left(\frac{\alpha_B}{\pi}\right)\right\} \cong
\frac{4}{b_0}\left(\frac{1}{\pi}arctg\left(\frac{\pi}{ln\frac{s}{\Lambda^2}}\right)+
\frac{2(M_0)^2}{s(ln\frac{s}{\Lambda^2})^3}+\ldots\right).
\label{133} \ee

This result can be most easily obtained from (\ref{129}), where
the contour of integration $C_R$ is modified to be a circle of
radius $R=s$ with center at $\sigma=0$, namely
\be
R(s)=\frac{1}{2i\pi}\int_{C_R}D(\sigma)\frac{d\sigma}{\sigma};~~
\Phi\left\{\left(\frac{\alpha_B}{\pi}\right)^k\right\}=\frac{1}{2i\pi}\int_{C_R}\frac{
d\sigma}{\sigma}\left(\frac{\alpha_B(\sigma)}{\pi}\right)^k
\label{134} \ee The result (\ref{132}) is obtained from
(\ref{134}) trivially introducing angular variable
$\sigma=s\exp(i\phi)$. For large $s,s\gg m^2,(M_0)^2$ one can use
in (\ref{134}) asymptotic form of $\alpha_B(\sigma)$ from
(\ref{110}).

Thus one obtains for  $R(s)$:
\be
R(s)=N_c\sum_q e^2_q\left [1+\bar  \Phi\left
\{\frac{\alpha_B}{\pi}\right\} + d_2\bar \Phi\left
\{\left(\frac{\alpha_B}{\pi}\right)^2\right\}+
d_3\bar\Phi\left\{\left(\frac{\alpha_B}{\pi}\right)^3\right\}+...\right]
\label{135}\ee Comparing with $
\Phi\left\{\left(\frac{\alpha_s}{\pi}\right)^k\right\}$ from
Radyushkin \cite{9}, one can see very close correspondence with
our $\bar\Phi\left\{\left(\frac{\alpha_B}{\pi}\right)^k\right\}$.

\section{Discussion and Conclusions}

We have developed in this paper the background perturbation theory
(BPT) in the large $N_c$ limit, when all physical amplitudes and
their perturbative expansions have only isolated singularities
(poles).

Using this fact, we have proposed a nonperturbative solution to
$\beta(\alpha)$ and $\alpha(Q^2)$ containing all terms  of loop
expansion.

By choosing a specific meromorphic  function we have modelled both
$\beta(\alpha_B)$ and $\alpha_B(Q^2)$ satisfying necessary
criteria of analyticity of $\alpha_B$ in the Euclidean region of
$Q^2$, and analyticity of $\beta(\alpha)$ for positive values of
$\alpha_B$.

We have constructed the time-dependent old-fashioned perturbation
theory  to investigate  analytic properties of different terms of
perturbation series, and have shown that those reduce to isolated
poles (of first and higher degree) shifted with respect to
unperturbed bound state positions.

This enables us to modify original ansatz for $\beta(\alpha)$ and
$\alpha_B(Q^2)$ to achieve a full correspondence with standard
perturbative expansion of $\beta(\alpha)$.

Finally, we have studied the problem of analyticity of
$\alpha_B(Q^2)$ and of analytic continuation into region of
time-like $Q^2$, where $\alpha_B(Q^2)$ has singularities, and have
compared this procedure with analytic continuation in SPT.

We have found a striking  similarity in the perturbative series
for $R(s)$ in case of SPT and BPT at large $s$. However our BPT
series has correct physical singularities at large $N_c$ in
contrast to SPT.

More detailed  calculations and comparison with experiment may
decide on the applicability of both methods and are planned for
the subsequent publication.

Financial support of the RFFI grant 00-02-17836 and INTAS grants
00-110 and 00366 is gratefully acknowledged.

\newpage

\section*{Appendix}

\setcounter{equation}{0} \def\theequation{A.\arabic{equation}}

 \vspace{1cm}

To find poles of $\alpha_B(s)=\frac{4\pi}{b_0\left[\ln\frac{
m^2}{\Lambda^2}+\psi\left(\frac{M^2_0-s}{m^2}\right)\right]}$  one
must find the roots $z_0(n)$ of the equation
\be
\ln\frac{m^2}{\Lambda^2}+\psi(z_0)=0\label{A.1}\ee which can be
represented as
\be
z_0(n)=-n+\delta,~~ \delta\ll 1,~~n=0,1,2.\label{A.2}\ee The
following relations will be useful
\be
\psi(-n+\delta)=\sum^n_{k=0}\frac{1}{k-\delta}+\psi(1+\delta),~~n\geq
0,\label{A.3}\ee
\be
\psi(1+\delta)=-C+\delta\{\zeta(2)-\delta\zeta(3)+...\}\label{A.4}
\ee where $\zeta(2)=\frac{\pi^2}{6},~~\zeta(3)=1.202$.

Hence one can write an expansion in powers of $\delta$ \be
\psi(-n+\delta)=-\frac{1}{\delta}-C+S_1(n)+\delta(1+S_2(n)+\zeta(2))+O(\delta^2)\label{A.5}\ee
where $S_k(n)=\sum^n_{k=1}\frac{1}{k},~~ C=0.577$, and
Eq.(\ref{A.1}) takes the form
\be
-\frac{1}{\delta}-C+S_1(n)+\delta(1+S_2(n)+\zeta(2))=-\ln\frac{m^2}{\Lambda^2}.\label{A.6}\ee
To the lowest order one has
\be
\delta\cong \delta_0=\frac{1}{\ln\frac{m^2}{\Lambda^2}-C+S_1(n)}
+O(\delta_0^3).\label{A.7}\ee

In a similar way one obtains $\psi'(z_0),$
\be
\psi'(z_0)=\frac{1}{\delta^2}+S_2(n)+\zeta(2)+O(\delta).\label{A.8}\ee
For $n\gg1, S_2(n)\approx\zeta(2)=\frac{\pi^2}{6}$.

In this way $\alpha_B$ in the vicinity of the pole is
\be
\alpha_B(s\approx
M^2_n+i\varepsilon)=\frac{4\pi}{b_0[\delta_0^{-2}(n)+S_2(n)+\zeta(2)](M^2_n-\delta_0m^2
-s)}\label{A.9}\ee
where $M^2_n=M^2_0+nm^2$.

Now for $S_1(n)$ the following asymptotics holds true
\be
S_1(n)=C+\ln
n+\frac{1}{2n}+O\left(\frac{1}{n^2}\right).\label{A.10}\ee
Therefore $\delta_0 $ (\ref{A.7}) can be rewritten as
\be
\delta_0\cong \frac{1}{\ln\frac{m^2}{\Lambda^2}+\ln
(n+\frac{1}{2})}=\frac{1}{\ln\frac{m^2(n+\frac{1}{2})}{\Lambda^2}}
\cong\frac{1}{\ln\frac{M^2_n}{\Lambda^2}}
\label{A.11}\ee and the averaged imaginary part of $\alpha_B$ in
(\ref{126}) has the form
\be
\lan Im\alpha_B\ran_{Nm^2} =\frac{4\pi^2}{b_0N}
\sum^n_{k=0}\frac{1}{\left(\ln\frac{M^2_{n+k}}{\Lambda^2}\right)^2+\frac{\pi^2}{3}}
\approx\frac{4\pi^2}{b_0\left(\ln\frac{\overline{M}^2}{\Lambda^2}\right)^2},\label{A.12}\ee
where $\overline{M}^2\sim s$ is in the middle of the averaging
interval, which is asymptotically close to the expression in Eq.
(\ref{125}).

\end{document}